\documentclass[12pt]{article}
\usepackage{amsmath}
\usepackage{amsthm}
\usepackage{amssymb}
\usepackage{amsfonts}
\usepackage{epic}
\usepackage{eepic}
 \usepackage{times}
\usepackage[matrix,arrow]{xy}

\usepackage{macros}
\usepackage{color}
\usepackage{framed}

\usepackage{epsfig} 
\definecolor{shadecolor}{rgb}{1,0.9,0.7}

\theoremstyle{plain}

\newtheorem{claim}{Claim}
\newtheorem{propn}{Proposition}
\newtheorem{lem}{Lemma}

\newtheorem{conj}{Conjecture}

\newtheorem{defn}{Definition}

\theoremstyle{remark}
\newtheorem{rem}{Remark}

\newcommand{\sst}{\scriptscriptstyle}

\newcommand{\vt}{\vartheta}
\newcommand{\ka}{\kappa}
\newcommand{\nn}{\nonumber}

\renewcommand{\1}{\one}
\renewcommand{\2}{\two}

\newcommand{\beq}{\begin{equation}}
\newcommand{\eeq}{\end{equation}}

\newcommand{\pa}{\partial}
\newcommand{\ot}{\otimes}
\newcommand{\ra}{\to}

\renewcommand{\L}{{\rm\sst L}}

\newcommand{\fr}[2]{{\textstyle \frac{#1}{#2} }}

\newcommand{\ket}{\rangle}

\newcommand{\al}{\alpha}

\newcommand{\ga}{\gamma}
\newcommand{\Ga}{\Gamma}
\newcommand{\de}{\delta}

\newcommand{\ep}{\epsilon}

\newcommand{\Om}{\Omega}
\newcommand{\si}{\sigma}

\newcommand{\vf}{\varphi}

\newcommand{\bk}{{\mathbf k}}
\newcommand{\by}{{\mathbf y}}

\newcommand{\CC}{{\mathcal C}}

\newcommand{\CF}{{\mathcal F}}

\newcommand{\CH}{{\mathcal H}}

\newcommand{\bfx}{{\mathbf x}}
\newcommand{\CI}{{\mathcal I}}

\newcommand{\CO}{{\mathcal O}}  
\newcommand{\CP}{{\mathcal P}}  
\newcommand{\CQ}{{\mathcal Q}}

\newcommand{\CY}{{\mathcal Y}}

\newcommand{\SA}{{\mathsf A}}
\newcommand{\SB}{{\mathsf B}}
\newcommand{\SC}{{\mathsf C}}
\newcommand{\SD}{{\mathsf D}}

\newcommand{\SH}{{\mathsf H}}

\newcommand{\SM}{{\mathsf M}}

\newcommand{\ST}{{\mathsf T}}

\newcommand{\SRN}{{\rm N}}
\newcommand{\SRM}{{\rm M}}

\newcommand{\FC}{{\mathfrak C}}

\newcommand{\FF}{{\mathfrak F}}

\newcommand{\0}{{\mathfrak 0}}
\newcommand{\one}{{\mathfrak 1}}
\newcommand{\two}{{\mathfrak 2}}

\newcommand{\BR}{{\mathbb R}}

\newcommand{\BC}{{\mathbb C}}

\newcommand{\BS}{{\mathbb S}}

\newcommand{\BT}{{\mathbf t}}

\newcommand{\BK}{{\mathbb K}}
\newcommand{\BZ}{{\mathbf Z}}
\newcommand{\BBZ}{{\mathbb Z}}

\DeclareMathOperator{\sgn}{sgn}

\newcommand{\rf}[1]{(\ref{#1})}

\newcommand{\aufz}
{\begin{list}{$\bullet$}{\topsep0cm \itemsep0cm \parsep0cm}}
\newcommand{\eaufz}{\end{list}}

\begin{document}


\title{On the spectrum of the Sinh-Gordon model in finite volume\footnote{This version, posted in 2017,
corrects a mistake in the lattice TBA of the previous versions.}}

\author{J. Teschner}

\address{DESY Theory,\\
Notkestr. 85\\
22603 Hamburg\\
Germany}

\begin{abstract}
We discuss the spectrum of the Sinh-Gordon model in finite volume.
\end{abstract}
\maketitle

\section{Introduction}

We will discuss  the quantization of the classical field theory 
which has the following Hamiltonian:
\begin{equation}\label{Hdens}
H_{\rm\sst ShG}\,=\,\int\limits_0^{R}\! dx\,
\left\{ 4\pi \,\Pi^2+\frac{1}{16\pi}(\pa_\si \vf)^2+2\mu\cosh(b\vf)
\right\}\,.
\end{equation}
Part of our motivation is the possible role of this model as a prototype
for integrable models with non-compact target spaces like the
nonlinear sigma models for non-compact symmetric spaces.

We will mainly consider the Sinh-Gordon model in finite spacial 
volume which corresponds to imposing the periodic boundary conditions
$\vf(x+R)=\vf(x)$. A lot is known about the Sinh-Gordon model
in infinite volume $R\ra\infty$, see e.g. \cite{VG,FMS,KMu,BL,Le}. 
The spectrum contains a single particle with mass $m$ which is related to 
the parameter $\mu$ in \rf{Hdens} in a definite way \cite{Z3}.
The S-matrix factorizes into the product of two-particle scattering phase
shifts $S(\vt_1-\vt_2)$, with
\begin{equation}
S(\vt)\,=\,\frac{\sinh\vt-i\sin\vt_0}
{\sinh\vt+i\sin\vt_0}\,,
\qquad \vt_0\,\equiv \frac{\pi b^2}{1+b^2}\,.
\end{equation} 
The known form-factors \cite{FMS,KMu,BL} completely characterize the
corresponding local fields of the theory, at least in principle. 

However, less is known about
the case of finite $R$. Basically nothing is known about the
spectrum of the model beyond certain characteristics
of the ground state \cite{Za1,Za2,Lu}. However, among these
papers we would like to mention in particular \cite{Za2} and 
\cite{Lu} as important precursors of our investigation. 
The paper \cite{Za2} discusses certain functional relations
related to 
the thermodynamic Bethe ansatz equations which include the 
Baxter equation and the so-called quantum 
Wronskian relation. These functional relations will
also play an important role in this paper.
In \cite{Lu} it was suggested 
that the separation of variables method (SOV-method) 
introduced by Sklyanin
should be useful for this problem. 
Certain conjectures were proposed which would fully
characterize the ground state expectation values of 
the fields $e^{\al\vf}$. These conjectures were
shown in \cite{Lu} to pass highly nontrivial tests.

In order to get a solid groundwork for solving the Sinh-Gordon model
we have in joint work \cite{BT} with A. Bytsko introduced
an integrable lattice regularization of the Sinh-Gordon model.
The construction of the Q-operator and the application of the 
SOV-method allowed us to characterize the spectrum of the
model by identifying the precise conditions which ensure that
a solution of the Baxter equation corresponds to a state in the
spectrum.  

In the present paper, which heavily builds upon 
\cite{BT} we will present nontrivial evidence in favor of the following 
claim. In order to formulate it, we will use the 
notation $\BT=[\vt_1,\dots,\vt_M]$, 
$\bk=[k_1, \dots,k_M]$ for vectors modulo permutations
of their components \footnote{This means that
$[\dots,k_a,\dots,k_b,\dots]=
 [\dots,k_b,\dots,k_a,\dots]$ with $k_a=k_b$ being allowed.}.


\begin{shaded}
\begin{claim} \label{mainclaim}
The Hilbert space $\CH_{\rm\sst SG}$ of the Sinh-Gordon model contains a 
sector $\CH_{\rm\sst TBA}$ which exists for all $R>0$ 
and is complete within $\CH_{\rm\sst SG}$
both in the infrared (IR) limit
$R\ra\infty$ and the 
ultraviolet (UV) limit $R\ra 0$, respectively.

$\CH_{\rm\sst TBA}$ 
decomposes 
into subspaces 
$\CH_M$ as $\CH_{\rm\sst TBA}=\oplus_{M=0}^{\infty}\CH_M$.
The sectors $\CH_M$ have an orthonormal basis 
spanned by eigenvectors $e_{\bk}$ to all the
conserved quantities of the Sinh-Gordon model which are labeled by 
tuples
$\bk=[k_1,\dots,k_M]\in \BBZ^M$ of integers. 
The eigenvalue $E_\bk$ of the Hamiltonian 
in the eigenstate $e_\bk$ can then be expressed as
\begin{equation}\label{Ek}
E_\bk\,=\, \sum_{a=1}^M m\cosh\vt_a -{m}\int_{\BR}
\frac{d\vartheta}{\pi}\,\cosh\vt \,  \log(1+Y_{\BT}(\vartheta)) 
\,,
\end{equation}
where the function $Y_{\BT}(\vt)$ and the tuple 
$\BT=[\vartheta_1,\dots,\vartheta_M]$
of real numbers are defined as follows.

\begin{itemize}
\item[(i)] For each tuple $\BT$ of real
numbers one may define $Y_{\BT}(u)$ 
as the unique solution
to the equations
\[
{\rm (I)}\qquad
\begin{aligned}
\log Y_\BT(\vartheta)-\int_{\BR}
\frac{d\vartheta'}{2\pi}\; \si(\vartheta & -\vartheta')  \log(1+Y_{\BT}
(\vartheta'))\\[-1ex] & \quad +mR\cosh\vartheta
 +\sum_{a=1}^M \log S(\vartheta -\vartheta_a-i\fr{\pi}{2})\,=0
 \,,
\end{aligned}
\]
where  the kernel $\si(\vt)$ is related to the $S(\vt)$ via
$
\si(\vt)\equiv\frac{d}{d\vt}\arg S(\vt).
$

\item[(ii)] There is a unique solution  $\BT=\BT(\bk;R)$ to the system
of equations
\[
{\rm (B)}\qquad
\begin{aligned}
mR\sinh\vartheta_a
+\sum_{\substack{b=1\\ b\neq a}}^M \arg & S(\vartheta_a-\vartheta_b)\\[-2ex] 
& +i\int_{\BR}\frac{d\vartheta}{2\pi}\,
\si(\vartheta_a-\vartheta+i\fr{\pi}{2}) \log(1+Y_{\BT}(\vartheta))\,=\,
2\pi k_a\,,
\end{aligned}
\]
such that the leading asymptotics of the tuples $\BT(\bk;R)$ for $R\ra\infty$
is given by the tuples $\BT_0(\bk)$ which are the 
unique solutions to the equations obtained from 
${\rm (B)}$ by setting $Y_\BT(\vt)\equiv 0$.
\end{itemize}
\end{claim}
\end{shaded}
In order to support our claim we will proceed as follows.
\begin{itemize}
\item In Section 2 we will
rigorously derive the lattice counterparts 
(I)$_\L$ and (B)$_\L$ of equations (I) and (B) from the
results of \cite{BT}.

\item It is then observed in Section 3 
that the continuum limit of these
equations yields minor generalizations of the
equations (I) and (B) above. We identify the 
conditions on the solutions of these equations which
describe the spectrum of the Sinh-Gordon model.
 
\item In Section 4 we show that $\CH_{\rm\sst TBA}$ is {\it complete}
in $\CH_{\rm\sst SG}$ in the infrared (IR) limit
$R\ra\infty$.

\item Section 5 discusses the UV asymptotics $R\ra 0$. 
Comparing the description of $\CH_{\rm\sst TBA}$ given above to 
the predictions which follow from 
the expected relation with 
Liouville theory we find nontrivial evidence for the 
claim that $\CH_{\rm\sst SG}\simeq \CH_{\rm\sst TBA}$ in the UV.
\end{itemize}

Given that $\CH_{\rm\sst TBA}$ appears to be {\it complete}
in $\CH_{\rm\sst SG}$ both in the UV and the IR, it is tempting
to speculate that the following conjecture might be true.

\begin{conj}\label{complconj}
We have $\CH_{\rm\sst SG}=\CH_{\rm\sst TBA}$ for all values of $R$.
\end{conj}


It is worth noting that the special case 
where the particle number $M$ is zero reproduces the 
thermodynamic Bethe ansatz (TBA) equation for the 
Sinh-Gordon model \cite{Za1}. Our proposal may therefore be 
seen as describing the generalization of the TBA to 
arbitrary excited states.

For $M\neq 0$ one may view the equations (B) 
as quantization conditions similar to the Bethe ansatz equations
which determine the spectrum. 
The term 
in (B) which contains $Y_\BT(\vt)$ 
represents finite volume corrections which 
are exponentially small when $R\ra\infty$. The leading behavior 
for $R\ra\infty$ is then precisely what one expects on the basis
of the fact that the infinite volume S-matrix factorizes into 
two-particle scattering phases, see section \ref{AsBethe} for 
more discussion.

Although the representation above is not fully
explicit, it is efficient in the following sense.
First note that it will certainly be possible to solve the
equations (I) and (B) numerically with the help of a 
computer. Compared to any direct approach to the 
diagonalization of the Hamiltonian (without
exploiting integrability) we have reduced the 
computational complexity enormously. 
Secondly, the system of equations (I), (B) allows us to 
extract a lot of important information like information
about the asymptotic behavior in the infrared $mR\ra \infty$
or ultraviolet $mR\ra 0$, respectively.

We should also admit that the discussion of the UV 
asymptotics that we can offer in Section 5 is neither 
fully rigorous nor complete. However, some of our observations appear 
to be fairly nontrivial and therefore worth being mentioned. 
In particular, by using certain results from \cite{FZ} we are able 
to recover the Liouville reflection amplitude
directly from the Sinh-Gordon model, and to 
give an alternative derivation 
of the important result from \cite{Z3} concerning the exact relation 
between the scale parameters defined in terms of the
UV and the IR behavior of the Sinh-Gordon model, respectively.


\section{Lattice Sinh-Gordon model}
\setcounter{equation}{0}

We briefly recall from \cite{BT} the definition of the lattice Sinh-Gordon 
model followed by the description of its spectrum that was found in this
reference.

\subsection{Definition of the lattice Sinh-Gordon model}

\subsubsection{Hilbert space $\CH$}

The Hilbert space $\CH$ is defined as 
\begin{equation}
\CH\,=\,(L^2(\BR))^{\SRN}\,.
\end{equation}
It is defined uniquely up to unitary equivalence as the irreducible 
representation of the canonical commutation relations
\[
{[}\,\Pi_n\,,\,\vf_n\,{]}\,=\,\frac{1}{2\pi i}\,\de_{n,m}\,.
\]
by self-adjoint operators $\vf_n$, $\Pi_m$, $n,m=1,\dots,\SRN$. 
The operators $\vf_n$ and $\Pi_n$
represent the discretized Sinh-Gordon field and 
its canonical conjugate momentum, respectively.

\subsubsection{Conserved charges}

Let us define the monodromy matrix $\SM$ as
\begin{equation*}\label{Mono}
 \SM(u) \,\equiv\, \left(\begin{matrix} \SA(u) & \SB(u) \\
 \SC(u) & \SD(u) \end{matrix}
 \right)\,\equiv\,L_{\sst\SRN}(u) \cdot \ldots \cdot L_\2(u) \cdot L_\1(u) \,,
\end{equation*}
where the Lax-operator $L_n(u)$ is chosen as 
\begin{equation*}\label{LL}
 L_n(u)  \,\equiv\,\frac{1}{i} 
 \left( \begin{array}{cc}  
  i\,e^{\pi b\Pi_n} \,
	\bigl( 1 +e^{-2\pi b (\vf_n+s)} \bigr) \,
	e^{\pi b\Pi_n} & 
e^{-\pi b s}\sinh \pi b (u + \vf_n)\\
e^{-\pi b s}\sinh \pi b (u - \vf_n)  & 
i\,e^{-\pi b\Pi_n} \,
	\bigl( 1 +e^{2\pi b (\vf_n-s)} \bigr) \,
	e^{-\pi b\Pi_n}
 \end{array} \right) \,,
\end{equation*}
Define the operators $\ST_m$ by the expansion
\begin{equation*}\label{Monotr}
 \ST(u) \,= \,\mathrm{tr}\, \bigl(\SM(u)\bigr)\,=\,e^{\pi b N u}\,
	\sum_{m=0}^{\SRN}(-e^{-2\pi b u})^m\,\ST_m,
\end{equation*}
The operators $\ST_n$, $n=1,\dots,\SRN$ are independent,
positive self--adjoint and mutually commuting. We
will denote the set $\{\ST_1,\dots,\ST_\SRN\}$ of conserved charges by 
$\FC$.

\subsubsection{Hamiltonian}

It was shown in \cite{BT} that there exists an operator $\SH$ which has
the following properties
\[
\begin{aligned}
{\rm (i)}\quad & \SH \;\;\text{is self-adjoint}.\\
{\rm (ii)}\quad & \SH \;\;
\text{commutes with $\ST_n$, $n=1,\dots,\SRN$}.\\
{\rm (iii)}\quad & \text{The classical continuum limit of $\SH$ yields 
\rf{Hdens}.}
\end{aligned}
\]
The explicit expression for $\SH$ is neither simple nor important for 
the following. We take properties (i)-(iii) as justification for 
calling the quantum theory defined by the triple of objects
\[
\big(\,\CH\,,\,\FC\,,\,\SH\,\big)
\]
the lattice Sinh-Gordon model.
The fact that the lattice Sinh-Gordon model
has equally many conserved charges as it has degrees
of freedom suggests that one may approach the spectral
problem for the Hamiltonian $\SH$ via the
spectral problem for the
conserved charges $\FC=\{\ST_1,\dots,\ST_\SRN\}$. 

\subsection{Characterization of the spectrum}

\subsubsection{Quantization conditions}

The main result in \cite{BT}, derived for odd $\SRN$, may be summarised as follows.

\begin{shaded}
\begin{claim}
A function $t(u)$
is a joint eigenvalue of the family of operators $\ST(u)$
if and only if it is $ib^{-1}$-periodic, $t(u)=t(u+i/b)$, 
and there exists 
a function $q(u)$ related to $t(u)$ by 
\begin{equation*}
{\rm (Bax)}
  \qquad t(u)q(u)\,=\,(a(u))^{\SRN}q(u-ib) +
(d(u))^{\SRN}q(u+ib)\,,
\end{equation*}
where $d(u)=a(-u)=1+e^{-2\pi b(u+s+i\frac{b}{2})}$, which has
the following properties,
\begin{equation*}
 \left[ \;\;
\begin{aligned}
{\rm (Q_1)} \quad & q(u)
\;{\sim} \;
\exp\big(-\mathrm{i}\fr{\pi }{2}(u\mp(s+i\de))^2 \SRN\big)\;\;\;{\rm for}\;\;
|u|\ra\infty,\;\,\; |{\rm  arg}(\pm u)|<\fr{\pi}{2} \,,\\
{\rm (Q_2)} \quad & q(u)\;\,
\text{is meromorphic with poles of maximal order $\SRN$ }
 \text{in}\;\,\pm\Upsilon_{-s},\\
& \text{the poles at $s\pm u={i}\de$ have order $\SRN$},
\end{aligned}
\;\;\right]
\end{equation*}
The set $\Upsilon_{s}$ is defined as $\Upsilon_{s}=s+{i}\de+
ib\BBZ^{\geq 0}+ib^{-1}\BBZ^{\geq 0}$ where $2\de\equiv b+b^{-1}$.
\end{claim}
\end{shaded}

It will also be useful to observe that the relevant functions
$q_t(u)$ can be characterized as solutions to a 
bilinear difference equation that only involves $q_t(u)$ itself.

\begin{propn}\label{wrprop}
$q(u)$ satisfies the 
so-called quantum Wronskian condition,
\begin{equation}\label{Wron}
  q(u+i\de) \, 
q(u-i\de)  -q(u+i\de') \, q(u-i\de')
\,=\,W_{\SRN}(u)\,,
\end{equation}
where we have used the notations 
$\de'\equiv \de-b$ and 
\begin{equation}
W_{\SRN}(u)\equiv \big[ e_b(-s+u-i\de)e_b(-s-u-i\de)\big]^{-\SRN},
\end{equation} with 
\begin{equation}
e_b(u):=\exp\left(\int_{\BR+i0}\frac{dt}{4t}\;\frac{e^{-2itu}}{\sinh(bt)\sinh(t/b)}
\right)\,.
\end{equation}
\end{propn} 
The fact that the functions $q_t(u)$ in the lattice Sinh-Gordon model
satisfy equation \rf{Wron}
was shown in \cite{BT}. 
There is also an
argument of F. Smirnov \cite{Sm} which may easily be 
adapted to deduce equation \rf{Wron} directly
from the conditions formulated in Claim 2.



\subsubsection{SOV representation}

It was furthermore shown in \cite{BT} that $\CH$ is unitarily equivalent to 
the space $\CH_{\rm\sst SOV}\equiv L^2(\BR^{\SRN},d\mu)^{\rm\sst Symm}$
which consists of functions $\Psi(\by)$, $\by=(y_1,\dots,y_{\SRN})$ 
that are square-integrable w.r.t. the measure
\begin{equation}
d\mu\,=\,\prod_{a=1}^{\SRN}dy_a\,\prod_{a>b}\,4\sinh\pi b (y_a-y_b)
\sinh \pi b^{-1}(y_a-y_b)\,,
\end{equation}
and totally symmetric under all permutations of the variables
$y_a$. The representation $\CH_{\rm\sst SOV}$
is essentially characterized by the following two
properties: 
\begin{itemize}
\item[(i)] The family of operators $\SB(u)$ is diagonal,
\[
\SB(u)\Psi(\by)\,=\,b_\by(u)\Psi(\by),
\]
where the eigenvalue $ b_\by(u)$ has been parametrized in terms of the
variables $y_a$ as
\[
b_\by(u)\,\equiv\,-\prod_{a=1}^{\SRN}\,e^{-\pi b s}\,2i\sinh\pi b (u-y_a)\,.
\]
\item[(ii)] Eigenstates $|\Psi_t\ket$ of $\ST(u)$ with eigenvalue $t(u)$
are represented by the wave-function
\begin{equation}\label{PsiSOV}
\Psi_t(\by)\,=\,\prod_{a=1}^{\SRN}q_t(y_a)\,,
\end{equation}
where $q_t(u)$ is an element of the set $\CQ$ define above.
\end{itemize}
It is worth keeping in mind that $q_t(u)$ not only represents the
eigenstates $|\Psi_t\ket$ via \rf{PsiSOV}, it also gives a generating
function for all the conserved charges of the theory.

\subsection{Reformulation in terms of integral equations}

The next step will be to reformulate the conditions characterizing the
spectrum in terms of nonlinear integral equations. The use of 
nonlinear integral equations as a reformulation of the functional 
relations that characterize the finite volume
spectrum of integrable models 
goes back to \cite{KP}.

Let $\BS$ be the strip
$\BS\equiv\{\,z\in\BC\,;\,|{\rm Im}(z)|\leq \de\,\}$.
We will then define $\CQ_M$ to be
the subset of $\CQ$ having $M$ zeros $u_a$, $a=1,\dots, M$ in $\BS$, counted with multiplicities.
It turns out that the parameters $u_a$ are particularly important
characteristics of the elements $q\in\CQ$.

\subsubsection{The integral equations}

Our aim will be to show that the elements of $\CQ_M$
are in one-to-one correspondence 
to the elements of a certain set $\CY_M$ 
of solutions to the following system of equations,
\[
{\rm (I)_\L}\qquad\quad 
\begin{aligned}
\log  Y(\vartheta) =& -\SRN\log\frac{\cosh\frac{\pi}{2\de}s+\cosh\vartheta}
             {\cosh\frac{\pi}{2\de}s-\cosh\vartheta}
-\,
\sum_{a=1}^{M} \log S(\vartheta-\vartheta_a-i\fr{\pi}{2})\\
&\qquad\quad+ \int_{{\BR}}
\frac{d\vartheta'}{2\pi}\;\si(\vartheta-\vartheta') \log(1+Y(\vartheta'))
\,,\\
\pi(2k_a+1)\,&=\,\int_{{\BR}}
\frac{d\vartheta}{2\pi}\;\tau(\vartheta_a-\vartheta) \log\left(1+{Y_{\mathbf u}(\vartheta)}\right)\\[-.5ex]
&\qquad+
\sum_{\substack{{b=1}\\{b\neq a}}}^{M} \arg S(\vartheta_a-\vartheta_b)
+2\SRN\arctan\left(\frac{\sinh\vartheta_a}{\cosh\frac{\pi}{2\de}s}\right)\,.
\end{aligned}
\]
In order to formulate the 
equations {\rm (I)$_\L$} 
we have been using the following definitions:
\begin{align}
&S(\vartheta)\,=\,
\frac{\tanh\frac{1}{2}(\vartheta-i\theta_0)}{\tanh\frac{1}{2}(\vartheta+i\theta_0)}\,,\qquad 
\theta_0=\frac{\pi b}{2\delta}\,,
\\
&\si(\vartheta)\,=\,\,\frac{\pa}{\pa \vartheta} \arg S(\vartheta)\,=\,
\frac{4\sin\theta_0\,\cosh\vartheta}
{\cosh2\vartheta-\cos2\theta_0}
\,,
\\
& \tau(\vartheta)=\mathrm{i}\si(\vartheta+i\de)=\frac{4\sin\theta_0\,\sinh\vartheta}
{\cosh2\vartheta+\cos2\theta_0}\,,\\
&\log\Xi_{\SRN}(\vartheta)\,=\,
\SRN\,\log\frac{\cosh\frac{\pi}{2\de}s+\cosh\vartheta}
             {\cosh\frac{\pi}{2\de}s-\cosh\vartheta}\,.
\end{align}

\subsubsection{The correspondence}

Our (first) main result is the following reformulation of the
conditions which describe the spectrum of the lattice Sinh-Gordon model:

\begin{shaded}
\begin{claim}\label{mthm} 
There is a one-to-one
correspondence between the solutions $Y(\vt)\in\CY_M$ of the integral
equations ${\rm (I)_\L}$ and the 
elements $q\in\CQ_M$. This  correspondence can be described as follows.

For a given element 
$q\in\CQ_M$ one gets the corresponding function $Y(\vt)$ via
\begin{equation}
1+Y(\vartheta)\,=\,Q(\vt+i\fr{\pi}{2})Q(\vt-i\fr{\pi}{2})\,,\qquad 
Q\big(\fr{\pi u}{2\delta}\big)\,=\,\Lambda_{\SRN}(u)q(u)\,,
\end{equation}
where 
$\Lambda_{\SRN}(u):=\big[\Phi_b(-s+u-i\de)\Phi_b(-s-u-i\de)\big]^{\SRN}$
with special function $\Phi_b(u)$ defined as
\begin{equation}
\Phi_b(u):=\exp\left(\int_{\BR+i0}\frac{dt}{8t}\;\frac{e^{-2itu}}{\sinh(bt)\sinh(t/b)
\cosh(2t\de)}\right)\,.
\end{equation}
The set $\BZ=\{\vt_1,\dots,\vt_M\}$ is the 
set of zeros of $Q(\vt)$ within $\BS$.

Conversely, given a solution 
$Y(\vt)\in\CY_M$  to the equations ${\rm (I)_\L}$
with $\vt_a\in\BS$  for $a=1,\dots,M$
one defines the corresponding 
element $q\in\CQ_M$ as
\begin{equation}\label{qreconst}
\begin{aligned}
\log\big(q\big(\fr{2\de}{\pi}\vartheta\big)\big)= -&\log\Lambda_{\SRN}(\vartheta)+
\sum_{a=1}^M\log\tanh\frac{1}{2}(\vartheta-\vartheta_a)\\
&+
\int_\BR\frac{d\vartheta'}{2\pi}\;\frac{1}{\cosh\frac{\pi}{2\de}(\vartheta-\vartheta')}
\log\left(1+{Y(\vartheta)}\right)\,.
\end{aligned}
\end{equation}
\end{claim}
\end{shaded}

The derivation can be found in Appendix A.

\section{Continuum Sinh Gordon model}
\setcounter{equation}{0}

\subsection{Continuum limit}

To begin with, we may observe that the equations
(I)$_\L$ have a well-defined limit when
$\SRN\ra\infty$, $\si\ra\infty$ such that
\begin{equation}\label{limdef}
{mR}\,\equiv\,
4\SRN e^{-\frac{\pi}{2\de}s}
\end{equation}
is kept constant.
The limit may be carried out on the level of the equations
(I)$_\L$  and leads to the system of equations
\[
 {\rm (I)}\qquad
\begin{aligned}
\log Y(\vartheta)& +mR\cosh\vartheta-\int_{{\CC}}
\frac{d\vartheta'}{2\pi}\;\si(\vartheta-\vartheta') \log(1+Y
(\vartheta'))\\
&\qquad -
\sum_{a=1}^{M'} \log S(\vartheta-\vartheta_a-i\fr{\pi}{2})\,=0\,,
\end{aligned} 
\]
together with the equations
\[
 {\rm (B)}\qquad
\begin{aligned}
\pi(2k_a+1)&=mR\sinh\vartheta_a+\int_{\CC}\frac{d\vartheta}{2\pi}\,
\tau(\vartheta_a-\vartheta) \log(1+Y(\vartheta))\\
&\qquad\qquad +\sum_{b=1}^{M} \arg S(\vartheta_a-\vartheta_b)
\,=0\,,
\end{aligned}
\] 
 restricting the positions of the points $\vartheta_a$, $a=1,\dots,M$.

\subsection{Reformulation in terms of the Baxter equation}

Working backwards we may reformulate the previous description
of the spectrum in terms of the Baxter equation.
To each function $Y(\vt)\in\CY_M$ we can associate
a function $q(u)$ via
\[
{\rm (Q)}\qquad
\begin{aligned}
\log q\big(\fr{2\de}{\pi}\vt\big)\,=\,&-{mR}\,
\frac{\cosh\vartheta}{2\sin\theta_0}+
\int_{\BR}\frac{d\vartheta'}{2\pi}\;
\frac{\log(1+Y(\vt'))}{\cosh(\vartheta-\vartheta')}
+\sum_{a=1}^M\log\tanh\frac{1}{2}(\vartheta-\vartheta_a) 
\,.
\end{aligned}
\]

We will regard the functions $q(u)$ that are defined in this way 
as the generating function for the conserved quantities of the continuum 
Sinh-Gordon model. The relation between the Q-operators in the lattice model
and in the continuum theory may involve a renormalisation described by
multiplication with an overall function of the spectral parameter, in general. The term proportional to
$mR$ in the definition of  $Q(\vartheta)$ above is motivated by the discussion in \cite{Lu}.

\begin{defn}
Let $\CQ$ be the set of all 
 solutions $q(u)$ of the functional equation
 \[ {\rm (Wr)} \qquad
q(u+i\de)q(u-i\de)-
q(u+i\de')q(u-i\de')\,=\, 1\,,
\]
which satisfy the conditions
\[
\left[
\begin{aligned}
{\rm (Q_1)}\quad  & q(u) \;\;\text{is entire analytic,}\\
{\rm (Q_2)}\quad  & \log q(u)\;\sim\;-
\frac{mR}{2\sin\vt_0}\cosh\frac{\pi}{2\de}u\;\;\text{for}\;\;
|{\rm Re}(u)|\ra\infty,\;\;|{\rm Im}(u)|<\de\,.
\end{aligned}
\right]
\]
\end{defn}

Arguments very similar to those used in the proof of Theorem \ref{mthm}
then lead us to the following claim.

\begin{claim}
Eigenstates
of the Hamiltonian of the continuum Sinh-Gordon model  
are in one-to-one correspondence
with the elements of $\CQ$.
\end{claim}

Let us note that $\CQ$ is a 
certain  set of solutions of Baxter's T-Q-relation
\begin{equation}
\label{TQ}
t(u)q(u)\,=\,q(u+ib)+q(u-ib)\,.
\end{equation}
Indeed, given $q\in\CQ$ let us use the notation 
$q_+(u)\equiv q(u)$, $q_-(u)\equiv q(u-i/b)$ and define
\begin{equation}
t(u)\,\equiv\,q_+(u+ib)q_-(u-ib)-
q_+(u-ib)q_-(u+ib)\,.
\end{equation}
It is then easy to verify that $t$ and $q$ satisfy the T-Q-relation
\rf{TQ} with $t(u)$ being $ib^{-1}$-periodic, $t(u+i/{b})=t(u)$.

\subsection{Discussion}

It seems reasonable to 
regard the conditions defining the set $\CQ$ as an abstract 
characterization of the spectrum of the Sinh-Gordon model. 
The zeros of $Q(\vt)$ contained in the strip $\BS$ play 
a distinguished role. 

In the following two sections we will present evidence that the
class of solutions for which all the zeros of $Q(\vt)$ are located on the
real line is of particular interest - this class of solutions will 
reproduce the particle picture in the IR and will allow us to make
contact with Liouville theory in the UV. It seems likely
that the states which correspond to this class of solutions 
are complete within $\CH_{\rm\sst SG}$. 

It is therefore worth pointing out that the problem to classify the 
relevant solutions to equations (I) and (B) for which $\vt_a\in\BR$ 
for all $a=1,\dots,M$ can be organized as follows. First note that
we have

\begin{propn}
There is a unique solution to equation ${\rm (I)}$ for any chosen 
tuple $\BT=[\vt_1,\dots,\vt_M]$ of real numbers.
\end{propn}
\begin{proof} The proof given in \cite{FKS} for  $M=0$ 
can easily be generalized to the case $M\neq 0$.
\end{proof}

We may therefore define a tuple of functions $\Phi_a(\BT)$ by 
$\Phi_a(\BT)=\int_{\BR}\frac{d\vartheta}{2\pi}\,
\tau(\vartheta_a-\vartheta) \log(1+Y_{\BT}(\vartheta))
$,  $a=1,\dots, M$, 
and regard the resulting equations ${\rm (\widehat{B})}$,
\[
{\rm (\widehat{B})}\qquad
mR\sinh\vartheta_a
+\sum_{\substack{b=1\\ b\neq a}}^M \arg  S(\vartheta_a-\vartheta_b)
+\Phi_a(\BT)\,=\,
2\pi k_a\,,
\]
as a system of equations of Bethe ansatz type.

It will be shown in the next section that for any given tuple
$\bk=[k_1,\dots,k_M]$ there
exists a unique solution to ${\rm (\widehat{B})}$ 
when $R$ is sufficiently large but still finite. It seems almost certain that
this solution can be continued to all values of $R$ by means of the 
equations ${\rm (\widehat{B})}$. Indeed, these 
equations can be used to derive equations which express
$\frac{d}{dR}\vt_a$ in terms of certain {\em real} functions of 
$\BT=[\vt_1,\dots,\vt_M]$.  
These equations do not exhibit singular behavior for 
any finite value of $R$, which makes us believe that the solutions
found in the IR limit can be uniquely continued to arbitrary values of $R$.
This prediction will be further supported by our discussion of the UV 
limit in section 5. We therefore have little doubt that it is correct.

The solutions $\BT(R)=[\vt_1(R),\dots,\vt_M(R)]$ of the 
equations (B) that are defined in this way
are uniquely determined by the tuples
$\bk=[k_1,\dots,k_M]$ of integers which appear as data in  
equations (B), i.e. $\BT(R)\equiv\BT(\bk;R)$.
We will use the notation $Y_\bk(\vt)$ and $Q_\bk(\vt)$ for the
corresponding $Y$- and $Q$-functions, in particular 
$Y_\bk(\vt)\equiv Y_{\BT(\bk;R)}(\vt)$.

In order to arrive at our main claim formulated in the introduction
it remains to accept the following conjecture.
\begin{conj}
The value $E_\bk$ of the energy in the state $e_\bk$ is given
by formula \rf{Ek}. 
\end{conj}

This conjecture is in particular supported by the 
observation that it reproduces the prediction for the 
ground state energy that is produced by the 
thermodynamic Bethe ansatz (TBA). The additional terms which 
appear in the case $M\neq 0$ represent the sum over the
particle energies that one finds in the IR limit, cf. 
Section \ref{AsBethe}.

\section{The IR limit}\label{IR}
\setcounter{equation}{0}

\subsection{Equation {\rm (I)}}

To begin with, let us make a few observation on the solutions to the
equations (I) in the limit $R\ra \infty$.
First, let us observe that the asymptotics of the solutions $Y(\vt)$ to
equation (I) for $|\vt|\ra\infty$ is always
determined by the ``source term''
\begin{equation}\label{Sdef}
J(\vt)\,\equiv\,\cosh\vartheta
 +\frac{1}{mR}\sum_{a=1}^{M'} \log S(\vartheta -\vartheta_a-i\fr{\pi}{2})\,,
\end{equation}
as $Y(\vt)\sim e^{-mR J(\vt)}$. 
This is seen by noting that the kernel $\si(\vt-\vt')$
of the integral is peaked around $\vt=\vt'$, decaying rapidly away
from this region. This implies that the asymptotics of the
integral in (I) for $|\vt|\ra\infty$ is determined by the 
asymptotics of $Y(\vt)$. Taking into account the property ${\rm (Y_1)}$
allows us to conclude that the contribution of the 
integral in equation (I) becomes small when $|\vt|\ra\infty$. 
This implies that $Y(\vt)\sim e^{-mR J(\vt)}$ for $|\vt|\ra\infty$.

To continue it will be  useful to regard equation (I) as a fixed point 
equation of the form $\SA\cdot D=D$ 
for the deviation $D(\vt)\equiv \log Y(\vt)+mRJ(\vt)$ of 
$\log Y(\vt)$ from its asymptotics $-mRJ(\vt)$.
The operator $\SA$ is the nonlinear integral
operator defined by 
\begin{equation}
\big(\SA\cdot D\big)(\vt)\,\equiv\,\int_{\BR}
\frac{d\vartheta'}{2\pi}\; 
\si(\vartheta  -\vartheta')  \log\big[1+e^{D(\vt')-mRJ(\vt')}\big]\,.
\end{equation}
We observe in particular that $\SA\cdot D=D$ implies that 
$D$ is small of order $e^{-mR}$.
We conclude that $Y$ is small of the order $e^{-mR}$ for $R\ra\infty$ 
up to corrections which are suppressed by further factors of $e^{-mR}$ .

\subsection{Equations ${\rm (B)}$}

Turning our attention to the equations ${\rm ( B)}$
we observe that $Y\sim e^{-mR}$ implies that the term which contains
$Y$ will for large $mR$ be much smaller than the other terms. 
We therefore feel tempted to conclude that solutions to the 
equations  ${\rm ( B)}$ will be in one-to-one correspondence to the
solutions of the equations which we obtain from ${\rm ( B)}$
by dropping the term which contains $Y$.
Let us therefore consider the following simplified system of 
equations:
\begin{equation}\label{logBethe}
\pi(2k_a-1)+mR\sinh\vartheta_a
+\sum_{b=1}^{M} \arg S(\vartheta_a-\vartheta_b)=0.
\end{equation}
It is sometimes also convenient to consider the exponentiated version
of these equations
\begin{equation}\label{asBethe}
e^{-imR\sinh\vartheta_a}\,=\,-\prod_{a=1}^{M} S(\vt_a-\vt_b)
\end{equation}

\begin{lem}
The solutions $\vt_a$, $a=1,\dots,M$ are real.
\end{lem}
\begin{proof}
We follow the idea from \cite{KI}. Note that 
\begin{equation}
|S(\vt)|^2\,=\,\frac{|\sinh\vt|^2+(\sin\vt_0)^2-2\sin\vt_0\cosh({\rm Re}(\vt))
\sin({\rm Im}(\vt))}{|\sinh\vt|^2+(\sin\vt_0)^2+2\sin\vt_0\cosh({\rm Re}(\vt))
\sin({\rm Im}(\vt))}\,,
\end{equation}
from which it is easily read off that 
\begin{equation}\label{Sabsval}\left\{
\begin{aligned} &|S(\vt)| \leq 1\\
&|S(\vt)| \geq 1 \end{aligned}\right\}\;\;
{\rm if}\;\;
\left\{
\begin{aligned}  0\leq  & \;{\rm Im}(\vt)\leq \pi\\
-\pi \leq  & \;{\rm Im}(\vt)\leq 0\,.
\end{aligned}\right\}
\end{equation}
Let $\vt_{\rm max}$ be the component of 
$\BT=[\vt_1,\dots,\vt_M]$ 
which has maximal imaginary part. By taking the absolute 
value of both sides of \rf{asBethe} it follows 
that ${\rm Im}(\vt_{\rm max})\leq 0$. 
Let $\vt_{\rm min}$ be the component of 
$\BT=[\vt_1,\dots,\vt_M]$ 
which has minimal imaginary part. In a way that is similar to the 
previous argument we infer that ${\rm Im}(\vt_{\rm min})\geq 0$. 
Taken together we have 
$0\leq {\rm Im}(\vt_{\rm min})\leq {\rm Im}(\vt_a)\leq 
{\rm Im}(\vt_{\rm max})\leq 0$, from which our claim follows.
\end{proof}

 We will  now consider the 
equations
\begin{equation}\label{logBethe2}
\pi(2k_a-1)+mR\sinh\vartheta_a
+\sum_{b=1}^M \arg S(\vartheta_a-\vartheta_b)=0.
\end{equation}

\begin{lem}\label{unilem}
For each choice of integers $\bk=[k_1,\dots,k_M]$ 
there is a unique solution $\BT=[\vt_1,\dots,\vt_M]$ to the 
equations \rf{logBethe2}.
\end{lem}

\begin{proof}
We follow the idea from \cite{YY}. Let us define 
\begin{equation}
P(\vt)\,\equiv\,\int_0^{\vt}d\vt'\;\arg S(\vt')\,.
\end{equation}
The equations \rf{logBethe2} are then easily seen to be conditions for 
a local extremum of the function
\begin{equation}
R(\BT)\,\equiv\,\sum_{a=1}^M (mR \cosh\vt_a+2\pi k_a\vt_a)+
\sum_{a<b}P(\vt_a-\vt_b).
\end{equation}
The matrix of second derivatives is positive definite,
\begin{equation}
\sum_{a,b}v_a\frac{\pa^2 R}{\pa\vt_a\pa\vt_b}v_b\,=\,
\sum_{a=1}^M v_a^2 \,mR \cosh\vt_a+\sum_{a<b}(v_a-v_b)^2\si(\vt_a-\vt_b)>0\,.
\end{equation}
It follows that the function $R(\BT)$ has a unique minimum, which
is the desired solution of \rf{logBethe2}.
\end{proof}

We would finally like to argue that existence and
uniqueness of the solutions to 
the equations ${\rm (\widehat{B})}$ will still hold if
$R$ is sufficiently large but finite. Indeed, the equations
${\rm (\widehat{B})}$ are the conditions for an extremum of  the function
\begin{align}\label{Rhatdef}
\widehat R(\BT)\,\equiv\,&\sum_{a=1}^M (mR \cosh\vt_a+2\pi k_a\vt_a)+
\sum_{a<b}P(\vt_a-\vt_b)\\
& +\int_{\BR}\frac{d\vt}{2\pi}\;F(Y_{\BT}(\vt))
-\int_{\BR}\frac{d\vt}{2\pi}\int_{\BR}\frac{d\vt'}{2\pi}\;
\log(1+Y_{\BT}(\vt))\,\si(\vt-\vt')\,\log(1+Y_{\BT}(\vt'))\,,
\nn\end{align}
where $F(y)\equiv \int_0^{y}\frac{dt}{t}\log(1+t)$.
The additional terms in \rf{Rhatdef} will be suppressed 
by factors of $e^{-mR}$.  For large enough values
of $R$ this is enough to conclude that 
their presence will not spoil positivity of the
matrix of second derivatives, which leads us to the conclusion
that existence and uniqueness of the solutions to 
${\rm (\widehat{B})}$ is ensured if $R$ is sufficiently large but still 
finite.

\begin{rem} In similar problems it is often assumed that $k_a\neq k_b$ 
if $a\neq b$. It should be stressed that this assumption is not 
coming from the existence of solutions to the 
equations \rf{logBethe2}. Instead it is coming from additional 
requirements like the existence of wave-functions of a certain form
associated to the solutions of
equations like \rf{logBethe2}. In our case we have shown that
all functions $Y(\vt)\in\CY_M$ correspond to states in the spectrum,
and we furthermore argued 
that for large enough $R$ 
there exist functions $Y_{\bk}(\vt)\in\CY_M$ corresponding to 
all tuples $\bk=[k_1,\dots,k_M]$, 
including those which may have $k_a=k_b$ for $a\neq b$. 
\end{rem}


\subsection{Relation to the particle picture}\label{AsBethe}

We have seen that to leading order in $e^{-mR}$ one gets the
spectrum from the solutions of the equations \rf{logBethe2}
of Bethe ansatz type.
This can be
understood more intuitively as follows.
The interactions between a pair of particles in a massive relativistic
quantum field theory like the Sinh-Gordon model are expected to 
be relevant mainly if the mutual distance is of the order $1/m$, with
rapid decay when the distance increases further.  
For small density $M/R$ it should therefore give a reasonable
approximation to assume that the motion of the particles is 
approximately free for most of the time, except if two particles 
cross each other. In this case one should be able to describe
the crossing of two particles in terms of the 
two-particle S-matrix. In order to get an approximate description
for a state $\Psi_\BT$ characterized by the rapidities
$\BT=[\vartheta_1,\dots,\vartheta_M]$ one may try
to represent it in terms of 
a coordinate space wave-functions $\Psi_{\BT}(\bfx)$,
${\mathbf x}=(x_1,\dots,x_M)$.  
The idea that the motion 
is free except for two-particle crossings represented
by the scattering phase shift $S(\vt_a-\vt_b)$ 
leads to the following
ansatz for the wave-function $\Psi_\BT(\bfx)$,
\begin{equation}
 \Psi_\BT^{}(\bfx)\,=\,\exp\bigg(i\sum_a^M p_ax_a\bigg)\sum_{p\in S_M}
K_{p}^{}(\BT)\chi_p(\bfx)\,,
\end{equation}
where $S_M$ is the set of all permutations $p$ of the elements
of $\{1,\dots,M\}$, $\chi_p(\bfx)=1$ if $x_{p(1)}<x_{p(2)}<\dots<x_{p(M)}$
and zero otherwise, and $K_{p}(\BT)$ is defined up to a 
multiplicative constant by the property that the element $(ab)\in S_M$
which generates an exchange
of the elements $a$ and $b$ of $\{1,\dots,M\}$ is realized as
\[
K_{(ab)p}(\BT)\,=\, 
S(\vartheta_a-\vartheta_b)K_{p}(\BT)\,.
\]
Requiring that the wave-function is periodic w.r.t. the particle 
coordinates $x_a$ when $x_a\ra x_a+R$ then leads to the quantization 
conditions 
\begin{equation}\label{AsympBethe}
e^{iRp_a}\prod_{\substack{b=1 \\ b\neq a}}^M 
S(\vartheta_a-\vartheta_b)\,=\,1\,.
\end{equation}
We see that  our Bethe ansatz equations (B) describe
finite volume deviations of order 
$e^{-mR}$ from the asymptotic Bethe ansatz equations
\rf{AsympBethe}.


\section{The ultraviolet limit}
\setcounter{equation}{0}

We will now see that our proposal for the spectrum of the Sinh-Gordon model
seems to lead to a rather nice picture of its UV limit.
Some qualitative and quantitative aspects of this picture
will be found to be in 
remarkable agreement with certain
predictions that follow from the
expected relationship with Liouville theory \cite{ZZ}.

\subsection{Partial decoupling of left- and right-movers}\label{UVspec1}

\subsubsection{Clustering of the roots of {\rm (B)}} \label{cluster}

We would like to argue that the solutions $\BT=[\vt_1,\dots,\vt_M]$ 
of equations {\rm (I)} and {\rm (B)} have the following behavior
for $R\ra 0$, 
\begin{equation}\label{UVroots}
 \vt_a \,=\,\sgn(k_a)|\log mR |+\CO(1)\,.\\
\end{equation}
This means that the particles with $\sgn(k_a)>0$  or $\sgn(k_a)<0$
are moving to the right  or to the left with velocity close to the 
speed of light, respectively.
The separations $|\vt_a-\vt_b|$ within each cluster are of the
order one in comparison with $|\log mR|$.

In order to see this, let us rewrite the equations (I), (B)
in terms of the variables $\theta_a$ and the functions
$\widetilde{Y}_{\bk}^\pm(\theta)$ which are defined respectively by
\begin{equation}\label{thetashift}
\vt_a\,=\,\theta_a+\sgn(k_a)|\log mR |,\qquad
\widetilde{Y}_{\bk}^\pm(\theta)\,=\,Y_\bk\big(\theta\pm|\log mR |\big)\,.
\end{equation}
Let $\BK_\pm$ be 
the sets of indices $a$ for which $\pm k_a\geq 0$ 
and let $\BK_0$ be 
the set of indices $a$ for which $k_a = 0$.
The equations (I), (B) can then be written in the following form: 
\[
\begin{aligned}
& {\rm (I'_\pm)}\qquad
\begin{aligned}
\log \widetilde Y_\bk^{\pm}(\theta)-\int_{\BR}
\frac{d\theta'}{2\pi}\; & \si(\theta  -\theta') 
\log(1+\widetilde Y_{\bk}^{\pm}
(\theta'))\\ & +\frac{1}{2}e^{\pm\theta}
 +\sum_{a\in\BK_\pm\cup\BK_0} 
\log S(\theta -\theta_a-i\fr{\pi}{2})+\CO(mR)\,=\,0
\,,
\end{aligned}
\\
&{\rm (B'_\pm)}\qquad
\begin{aligned}
\pi (2k_a-1)-\int_{\BR}\frac{d\theta}{2\pi}\;
& \tau(\theta_a-\theta)  \log(1+
\widetilde Y_{\bk}^{\pm}(\theta))
\\
& \pm\frac{1}{2}e^{\pm\theta_a}
+\sum_{b\in\BK_\pm\cup\BK_0} \arg  S(\theta_a-\theta_b)+\CO(mR)\,=\,0\,.
\end{aligned}
\end{aligned}
\]
We observe that the variables $\theta_a$ 
and the functions $\widetilde Y_{\bk}^{\pm}(\theta)$
are constrained by 
equations which up to corrections of order $mR$ do not 
depend on the scale parameter $mR$ at all.
The leading asymptotics of the functions $\widetilde Y_{\bk}^{\pm}(\theta)$
will therefore be found within the set of solutions to the equations
obtained from $({\rm I}_\pm')$, $({\rm B}_{\pm}')$ by dropping all
$mR$-dependent terms.

\subsubsection{The decoupled theories for left- and right-movers}
\label{masslessTBA}

Let us consider the equations obtained from the equations
$({\rm I}_\pm')$, $({\rm B}_{\pm}')$ 
above by dropping the terms of order $\CO(mR)$ and by
undoing the shifts \rf{thetashift} in order to re-introduce a convenient
$mR$-dependence. For reasons that
will become clear later we will also temporarily drop the terms
from the summations over $a\in\BK_\0$ and consider the equations
\[
\begin{aligned}
& {\rm (I_\pm)}\qquad
\begin{aligned}
\log Y^{\pm}(\vartheta)-\int_{\BR}
\frac{d\vartheta'}{2\pi}\; & \si(\vartheta  -\vartheta') 
\log(1+ Y^{\pm}
(\vartheta'))\\ & +\frac{ mR }{2}e^{\pm\vt}
 +\sum_{a\in\BK_\pm} \log S(\vartheta -\vartheta_a-i\fr{\pi}{2})\,=\,0
\,,
\end{aligned}
\\
&{\rm (B_\pm)}\qquad
\begin{aligned}
\pi (\pm 2k_a-1)-\int_{\BR}\frac{d\vartheta}{2\pi}\;
& \tau(\vartheta_a-\vartheta)  \log(1+
Y^{\pm}(\vartheta))
\\
& \pm\frac{ mR }{2}e^{\pm\vartheta_a}
+\sum_{b\in\BK_\pm} \arg  S(\vartheta_a-\vartheta_b)\,=\,0\,.
\end{aligned}
\end{aligned}
\]
The
corresponding $Q$-functions are defined respectively by
\[
{\rm (Q_\pm)}\quad
\log Q^\pm(\vartheta)=-
\frac{{mR}}{4\sin\vt_0}e^{\pm \vt}+
\int_{\BR}\frac{d\vartheta'}{2\pi}\; 
\frac{\log(1+Y_{\bk}^{\pm}(\vt'))}{\cosh(\vartheta-\vartheta')}
+ \sum_{a\in\BK_\pm}\log\tanh\frac{1}{2}(\vartheta-\vartheta_a)
\,.
\]
In a companion paper \cite{BT2} we will show that these 
equations characterize the spectrum of a quantum
integrable model that might be called
the $c>1$ quantum KdV theory\footnote{So far we had been assuming 
$b\in\BR$ which would correspond to $c>25$. 
The existence of an analytic continuation to 
$c>1$ is nontrivial. It ultimately follows from the
fact \cite{TR} that Liouville theory and, relatedly,
the theory of representations of the
Virasoro algebra have nice analytic properties
w.r.t. to the value of $c$.} in order to distinguish it from its
$c<1$ counterpart studied in \cite{BLZ0}. 

It is very important to notice that 
the system of equations ${\rm (I_\pm)}$, ${\rm (B_\pm)}$
has families of solutions that depend on an additional complex
parameter $P$. This is a new feature compared to the 
case of the equations ${\rm (I)}$, ${\rm (B)}$.
More precisely we have the following claim for which 
strong support will be given in \cite{BT2}.

\begin{claim} {\rm \cite{BT2}} \label{Qpmasym} For given
tuples $\bk=[k_1,\dots,k_{M_{\pm}}]$ 
of integers with $k_a>0$
and each imaginary value $P\in i\BR$
there exist solutions $\big(\,
Y_{\bk,P}^{\pm}(\vt)\,,\,\BT^{\pm}_{\bk,P}\,\big)$
of the equations $ {\rm (I_\pm)}$, ${\rm (B_\pm)}$,
respectively, uniquely specified by the
property that the 
corresponding functions $Q_{\bk,P}^{\pm}(\vt)$ have
leading asymptotics  for $\vt\ra\mp\infty$ of the 
following form
\begin{equation}\label{Ypmasym}
Q_{\bk,P}^{\pm}(\vt)\,\underset{\vt\ra\mp\infty}{\sim}\,
\frac{\cos(4\de P\vt\pm\Theta(\bk|P))}
{\sqrt{\sinh(2\pi bP)\sinh(2\pi b^{-1}P)}}\,.
\end{equation}
The 
function $\Theta(\bk|P)$, regarded as a function of $P$, 
are entire analytic and satisfy 
\[
\begin{aligned}
{\rm (a)} \quad & \text{$\Theta(\,\bk\,|\,P\,)$ is strictly monotonous.
}\\ 
{\rm (b)} \quad & \text{$\Theta(\,\bk\,|\,P\,)=-\,
\Theta(\,\bk\,|-P\,)$.}
\end{aligned}
\]
The corrections to the asymptotics \rf{Ypmasym} are suppressed by 
(non-integral) powers of $mR e^{\pm\vt}$.
\end{claim}

It follows from the last statement in this claim that
formula \rf{Ypmasym} may be expected to yield
reasonable approximations for the functions $Q_{\bk,P}^{\pm}(\vt)$
as long as $\pm\vt> -|\log mR|+\Lambda$  if $\Lambda\ll |\log mR|$ 
is such that
$e^{- \Lambda}\ll 1$.
When $mR$ is small one therefore expects to find a region
\begin{equation}\label{plateaudef}
\CP\,\equiv\,\big\{\,\vt\in\BR\,;\,|\vt|\,<\,|\log mR |-\Lambda\,\big\}\,,
\end{equation}
in which equation \rf{Ypmasym} provides  
good approximations for both  $Q_{\bk,P}^+(\vt)$
and $Q_{\bk,P}^-(\vt)$, respectively.
The region $\CP$ defined in \rf{plateaudef} will 
henceforth be called the plateau. 

We will in the following be interested in cases where $P\in\BR$
and $P=\CO(|\log(mR)|^{-1})$. Note that $Q_{\bk,P}^{\pm}(\vt)$ 
will then have infinitely many
additional zeros within the strip $\BS$,
as indicated by the asymptotics
\rf{Ypmasym}.\footnote{These remarks
clarify and correct a previous version of this paper.
The author is grateful to W. Nahm for a question which 
stimulated this discussion.} The corresponding 
functions $Y_{\bk,P}^\pm(\vt)$ will satisfy equations 
similar to ${\rm (I_\pm)}$, ${\rm (B_\pm)}$ which contain 
extra terms corresponding to the  additional zeros of 
$Q_{\bk,P}^{\pm}(\vt)$. These extra zeros are precisely
what is needed for describing the asymptotics of the  
solutions to $({\rm I}_\pm')$, $({\rm B}_{\pm}')$ 
in terms of the functions $Y_{\bk,P}^{\pm}(\vt)$
introduced in Claim \ref{Qpmasym}.

\subsubsection{Residual coupling}

We have observed in \S\ref{cluster} that the function 
$Y_{\bk}(\vt)$ we are ultimately interested
in has 
asymptotics for $mR\ra 0$ given by solutions to the equations 
$({\rm I}_\pm)$, $({\rm B}_{\pm})$. 
In \S\ref{masslessTBA} we introduced a certain class of solutions to these
equations. In order to proceed we will now assume that 
the asymptotics of $Y_{\bk}(\vt)$ for $mR\ra 0$ can be found within the
class of solutions to discussed in \S\ref{masslessTBA}. 
More precisely, we will assume that for $\bk_\pm$ chosen as
\begin{equation}\label{bkpmdef}
\bk_\pm\,=\,[\,\pm k_a\,|\,a\in\BK_\pm\,]\,.
\end{equation}
there will exist a distinguished 
value for $P$, $P\equiv P(\bk)$, such that
the functions $Y_{\bk_+,P(\bk)}^+(\vt)$
and $Y_{\bk_-,P(\bk)}^-(\vt)$ will both give good approximations to  
$Y_{\bk}(\vt)$ for values of $\vt$ on the plateau. 
This assumption can be supported substantially by comparison with
the lattice theory where an analogous statement can be 
checked explicitly \cite{BT2}.

We should keep in mind that
$-Q_{\bk_\pm,P}^\pm(\vt)$ and $Q_{\bk_\pm,P}^\pm(\vt)$ correspond to the 
same functions $Y_{\bk_\pm,P}^\pm(\vt)$, respectively. The requirement that
the asymptotics \rf{Ypmasym} provides a good approximation for the values
(up to a sign) that the function $Q_{\bk}(\vt)$ takes on the plateau 
then implies that $P(\bk)$ must be a solution to the condition
\begin{equation}\label{Q-cond}
{\quad
\Theta(\,\bk_+|\,P\,)+\Theta(\,\bk_-|\,P\,)\,\in\,
\pi\BBZ\,.\quad}
\end{equation}
We will label the different solutions $P$ of \rf{Q-cond}
by an integer $n$.
It follows from property (a) that $P_m<P_n$ if $m<n$.
Note furthermore that property (b) above 
implies that $P_{-n}=-P_n$. It follows that the solutions $P_n$
and $P_{-n}$ correspond to the same function
$Q(\vt)$. The solutions $P$ to the quantization conditions
are therefore uniuely determined by the data $(\bk_+,\bk_-,|n|)$.

Let us furthermore observe that the dependence of 
$\Theta(\,\bk_\pm|\,P\,)$ w.r.t. the variable $R$ is given by
\begin{equation}
\Theta(\,\bk_\pm|\,P\,)\,=\,\widehat{\Theta}(\,\bk_\pm|\,P\,)
+4\de P\,|\!\log mR|\,,
\end{equation}
where $\widehat{\Theta}(\,\bk_\pm|\,P\,)$ is independent of $R$. This 
follows from the simple fact that the dependence w.r.t. $mR$ can be removed
from the equations (I$_\pm$), (B$_\pm$) by a shift of the variable $\vt$.
This circumstance implies that the solutions to the quantization 
condition \rf{Q-cond} behave as 
\[
P_n\,\sim\,\frac{n\pi}{8\de|\log{mR}|}
\big(1+\CO((\log{mR})^{-1})\big)\,.
\]
For values of $n$ much smaller than $|\log{mR}|$ we may 
observe that the zeros of the function $Q_\bk(\vt)$ which 
fall into the plateau are separated by a distance of order
$|\log{mR}|$ from each other and from the boundaries
of the plateau. The number of these zeros is $|n|-1$.
Keeping in mind that the zeros $\vt_a$ of $Q_\bk(\vt)$ which do not 
correspond to the rapidities of the left- or rightmoving particles 
are 
those for which $k_a=0$ we are led to identify
\begin{equation}\label{nvsn0}
|n|-1\,=\,n_\0\,, 
\end{equation}
where $n_\0$ is the number of indices $a$ such that $k_a=0$.

Equations \rf{bkpmdef} and  
\rf{nvsn0} establish a one-to-one correspondence between the 
tuples $\bk$ which label the solutions $Y_\bk(\vt)$ of (I), (B)
and the data $(\bk_+,\bk_-,|n|)$ which label the 
solutions to the quantization conditions \rf{Q-cond}.
This means that there is a one-to-one correspondence 
between the vectors $e_\bk$ which span $\CH_{\rm\sst TBA}$ 
and the set of solutions  to the 
quantization conditions \rf{Q-cond}.

\subsubsection{Calculation of the energies}

Let us now observe that the eigenvalues $E_\bk$
of the Hamiltonian
can be calculated 
explicitly in terms of the 
data $(\bk_+,\bk_-,P_{n})$. 
\begin{propn}\label{confdim}
We have
\begin{equation}\label{UVenergy}
E_\bk\,=\,\frac{2\pi}{R}\bigg( 2P_{n}^2-\frac{1}{12}+\sum_{a\in \BK_+}
k_a+\sum_{a\in \BK_-}|k_a|\bigg)\,.
\end{equation}
\end{propn}
The proof of Proposition \ref{confdim} is given in Appendix B.

To summarize our conclusions: The part of the spectrum
contained in $\CH_{\rm\sst TBA}$ has UV asymptotics 
characterized by formula \rf{UVenergy} with 
$P_{n}\equiv P_{n}(\,r\,|\,\bk_+,\bk_-)$ being
defined by \rf{Q-cond}.

\subsubsection{Quantization conditions in the case $\BK_\pm=\emptyset$}

Let us consider the 
quantization conditions \rf{Q-cond}
for the case that $\BK_\pm=\emptyset$ where they simplify to
\begin{equation}\label{Q-red}
{\quad
2\Theta(P)\,\in\pi\BBZ\,.\quad}
\end{equation}
where $\Theta(P)\equiv \Theta(\,\bk_\pm|\,P\,)$ for $\BK_\pm=\emptyset$.

An explicit formula for the
functions  $\Theta(P)$ can be extracted
from the recent work \cite{FZ}. This works as follows.
A special case of the results of \cite{FZ} is a 
simple characterization of the functions $Q^{\pm}(\vt)$
for the case $\BK_\pm\!=\emptyset$ in terms of the solutions
to an ordinary differential 
equation, generalizing similar results for other models which
go back to \cite{DT,BLZ1}. The differential 
equation relevant for the present case is
\begin{equation}\label{ODE}
\left[-\frac{d^2}{dx^2}-p^2+\ka^2\big(e^{2x}+e^{-2x/b^2}\big)
\right]\Psi=0\,.
\end{equation}
There are unique solutions $\Psi_\pm$ to \rf{ODE} which have the
asymptotic behavior 
\begin{equation}\begin{aligned}
& \Psi_+\,\sim\,\frac{1}{\sqrt{2\ka}}\exp\left(
\frac{x}{2b^2}-\ka b^2 e^{-x/b^2}\right) \quad
{\rm for}\quad x\ra -\infty\,, \\
& \Psi_-\,\sim\,\frac{1}{\sqrt{2\ka}}\exp\left(-\frac{x}{2}-\ka e^{x}\right)
 \quad
{\rm for}\quad x\ra +\infty\,,
\end{aligned}
\end{equation}
respectively. The functions $Q^{\pm}(\vt)$ are then simply 
given as 
\begin{equation}
Q^+(\vt)\,\equiv\,Q^-(-\vt)\,\equiv\,
\Psi_+\frac{d}{dx}\Psi_--\Psi_-\frac{d}{dx}\Psi_+\,, 
\end{equation}
provided that we identify the respective parameters as
follows,\footnote{Concerning the comparison with \cite{FZ} let us note that 
the parameter $n$ used there is related 
to $b^2$ via $n=2/b^2$.}
\begin{equation}
\kappa\,=\,-\frac{\ka_0}{2\sin\frac{\pi b^2}{1+b^2}}\,\frac{mR}{2}\,e^{\vt},
\quad\ka_0\,=\,-
\frac{2\sqrt{\pi}}{\Ga\big(-\frac{1}{2(1+b^2)}\big)
\Ga\big(1-\frac{b^2}{2(1+b^2)}\big)}\,,\quad bp=2P\,.
\end{equation}
The characterization of $Q^{\pm}(\vt)$ in terms of the ODE \rf{ODE} 
allowed the authors of \cite{FZ} to 
determine the asymptotics of $Q^{\pm}(\vt)$. The explicit 
expression 
for $e^{2i\Theta(P)}$ which follows from
formula (177) in \cite{FZ}
is given by the formula
\begin{equation}\label{Thetadef}
e^{2i\Theta(P)}\,=\,-
\rho^{-8i\de P}\,
\frac{\Ga(1+2ibP)\Ga(1+2ib^{-1}P)}{\Ga(1-2ibP)\Ga(1-2ib^{-1}P)}\,,
\end{equation}
in which we have used the abbreviation
\begin{equation}
\label{rhodef}
\rho\,\equiv\, \frac{R}{2\pi}\,\frac{m}{4\sqrt{\pi}}\,
\Ga\bigg(\frac{1}{2+2b^2}\bigg)
\Ga\bigg(1+\frac{b^2}{2+2b^2}\bigg)\,.
\end{equation}
Having the explicit formula \rf{Thetadef} at hand, we may easily 
verify that the conditions (a), (b) 
as formulated in Claim \ref{Qpmasym} are satisfied for this case.

\subsection{Quantization conditions from Liouville theory}

The basic idea \cite{ZZ} is that the 
quantum Hamiltonian of the Sinh-Gordon model can be represented in the
following schematic form
\begin{equation}\label{HShG}
{\mathbf H}_{\rm ShG}\,=\,\frac{2\pi}{R}\int\limits_0^{2\pi}\! d\si\,
\left\{ 4\pi \,\Pi^2+\frac{1}{16\pi}(\pa_\si \vf)^2+\left(
\frac{R}{2\pi}\right)^{4b\de}\mu(e^{b\vf}+e^{-b\vf})
\right\}\,.
\end{equation}
For $R\ra 0$ one observes that in configuration space there is
a large region \begin{equation}\label{region}
-4\de\log\fr{2\pi}{R}+\fr{1}{b}\log\mu\ll\vf
\end{equation}  
where one can neglect the
interaction term $e^{-b\vf}$. 
Dropping this term yields the Hamiltonian ${\mathbf H}_{\rm L}$ 
of quantum 
Liouville theory. 

\subsubsection{The spectrum of Liouville theory}

Let us quickly review the relevant aspects of the quantum Liouville 
theory as constructed in \cite{TL}, see \cite{TR} for a review and
further references.
The spectrum of ${\mathbf H}_{\rm L}$
may be represented in the form
\begin{equation}\CH_{\rm L}\,=\,\int_0^{\infty}dP\;\CF^+_P\ot\CF^{-}_P\,,
\end{equation}
where $\CF^\pm_P\simeq \CF^\pm$ for all $P\in\BR_+$, with
$\CF^{\pm}$ being standard Fock spaces generated 
by  oscillators $a_{k}^{\pm}$, $[a_k^{\pm},a_l^{\pm}]=\frac{k}{2}\de_{k+l,0}$ 
from the Fock vacua $\Om^\pm$. The spaces  $\CF^{\pm}_P$ have bases
spanned by the vectors 
\begin{equation}
e_{P,\bk}^\pm\,=\,a_{-k_{M}}^{\pm}\dots a_{-k_1}^{\pm}\Om^{\pm},\qquad
\bk=[k_1,\dots,k_M]\,.
\end{equation}
The spaces $\CF^\pm_P$ carry an irreducible representation of 
two commuting copies of the Virasoro algebra with generators
\begin{equation}\begin{aligned}
& L_n^{\pm}(P)\,=\,2(P+in\de)a_n+\sum_{k\neq 0,n}a_k^{\pm} a_{n-k}^{\pm}\,,\\
& L_0^{\pm}(P)\,=\,P^2+\de^2+2\sum_{k>0} a_{-k}^{\pm} a_k^{\pm}\,.
\end{aligned}
\end{equation}
The central charge $c$ is given in terms of $\de$ by the formula
\begin{equation}
c\,=\,1+24\de^2\,.
\end{equation}
The action of 
the Hamiltonian ${\mathbf H}_{\rm L}$ of quantum Liouville theory
on $\CF^+_P\ot\CF^{-}_P$ is then represented by
\begin{equation}
{\mathbf H}_{\rm L}(P)\,=\,\frac{2\pi}{R}\,\bigg(
L_0^+(P)+L_0^-(P)-\frac{c}{12}\bigg)\,.
\end{equation}
which means that the states $e_{P,\bk_+}^+\ot e_{P,\bk_-}^-$ have eigenvalue
given by
\begin{equation}
{\mathbf H}_{\rm L}(P)\cdot
e_{P,\bk_+}^+\ot e_{P,\bk_-}^-\,=
\,\frac{2\pi}{R}\,
\Bigg(2P^2-\frac{1}{12}+\sum_{a=1}^{M_+}k_a^++\sum_{a=1}^{M_-} k_a^-\Bigg)\,
e_{P,\bk_+}^+\ot e_{P,\bk_-}^-\,.
\end{equation}

\subsubsection{The reflection operator}

For $R\ra 0$ there is a large  region in the configuration  space of
Liouville theory
\begin{equation}\label{region2}
\vf\ll 4\de\log\fr{2\pi}{R}-\fr{1}{b}\log\mu
\end{equation}
within which  one expects  that
it should be possible to describe the states 
of the Liouville theory in terms of a Schr\"odinger
representation for the zero mode $\vf_0=\int_0^{2\pi}
d\si \,\vf$, see \cite{TR} 
for a discussion of scope and limitations of such a representation.
Eigenstates $\Psi_P$ would then be represented 
by means of wave-functions $\Psi(\vf_0)\in\CF^+\ot\CF^{-}$ 
which have the form
\begin{equation}\label{Schrrep}
\Psi(\vf_0)\,=\,\left(\frac{R}{2\pi}\right)^{4i\de P}\!\!
e^{iP\vf_0}\, \FF_P\,+\,\left(\frac{R}{2\pi}\right)^{-4i\de P}\!\!
e^{-iP\vf_0}\,
\big({\mathbf T}(P)\cdot\FF_P\big)\,,
\end{equation}
where $\FF_P\in\CF^+\ot\CF^{-}$, and ${\mathbf T}(P)$ is the 
so-called reflection operator which describes the
reflection by the Liouville interaction
$e^{b\vf}$ which remains in \rf{HShG} 
after one has neglected $e^{-b\vf}$ \cite{ZZ,TR}. 

The operator 
${\mathbf T}(P):\CF^+\ot \CF^-\ra \CF^+\ot \CF^-$ is fully characterized
by the following properties:
\begin{itemize}
\item The operator ${\mathbf T}(P)$ can be factorized as
\[
{\mathbf T}(P)\,=\,T(P)\;{\mathbf T}^{+}(P)\ot{\mathbf T}^{-}(P)\,.
\]
\item The operators ${\mathbf T}^{\pm}(P)$
are completely defined by the 
intertwining property
\begin{equation}\label{intertw}
 {\mathbf T}^{\pm}(P)\cdot L_n^{\pm}(P)\,=\,
L_n^{\pm}(-P)\cdot{\mathbf T}^{\pm}(P)\,,
\end{equation}
together with the requirements that
\begin{equation}\begin{aligned}
&({\mathbf T}^{+}(P)\ot 1)\cdot(\Omega^+\!\ot\Omega^-)\,=\,
(\Omega^+\!\ot\Omega^-)\,,\\
&(1\ot {\mathbf T}^{-}(P))\cdot(\Omega^+\!\ot\Omega^-)\,=\,
(\Omega^+\!\ot\Omega^-)\,.
\end{aligned}\end{equation}
\item
The reflection amplitude $T(P)$ is given by the formula \cite{DO,ZZ}
\begin{equation}\label{refamp}
T(P)\,=\,-
\big(\pi \mu\ga(b^2)\big)^{-2iP/b}\,
\frac{\Ga(1+2ibP)\Ga(1+2ib^{-1}P)}{\Ga(1-2ibP)\Ga(1-2ib^{-1}P)}\,.
\end{equation}
\end{itemize}

Note in particular that 
that the unitary operators
${\mathbf T}^{\pm}(P)$ preserve the subspaces of $\CF^{\pm}$
defined by $\sum_{a=1}^{M_\pm} k_a^\pm={\rm const.}$. We may therefore
introduce eigenstates ${\mathfrak f}^{\pm}(\,{\bk_\pm}\,|\,P\,)\in\CF^{\pm}$ 
of ${\mathbf T}^{\pm}(P)$
which will form bases for $\CF^{\pm}$ when $\bk_\pm$ vary over all
tuples $[k_1^\pm,\dots,k_{M_\pm}^\pm]$ of integers with $k_a^{\pm}>0$.
The corresponding eigenvalues will be 
written as $\exp(i\theta_\pm (\,{\bk_\pm}\,|\,P\,))$.

\subsubsection{Quantization conditions of the Sinh-Gordon model in the UV}

In the case of the Sinh-Gordon model one would expect that 
a representation for the eigenstates of the Hamiltonian
in the form \rf{Schrrep} should be possible in the 
region 
\begin{equation}\label{region3}
-4\de\log\fr{2\pi}{R}+\fr{1}{b}\log\mu\ll
\vf\ll 4\de\log\fr{2\pi}{R}-\fr{1}{b}\log\mu\,.
\end{equation}
However, as opposed to the Liouville case one may no longer
choose $P$ arbitrarily. 
One way to find the restrictions on $P$ is to 
observe that the wave-functions 
of energy eigenstates in the 
parity symmetric potential $\cosh(b\vf)$ should have definite
parity $\pm 1$. The ground-state, in particular, should have positive
parity. Imposing this requirement on \rf{Schrrep} with $\FF_P$ 
being chosen as ${\mathfrak f}^{+}(\,{\bk_+}\,|\,P\,)\ot
{\mathfrak f}^{-}(\,{\bk_-}\,|\,P\,)$
leads to the quantization 
conditions
\begin{equation}\label{Liouquant}
\arg{T(P)}+\theta_+ (\,{\bk_+}\,|\,P\,)+
\theta_- (\,{\bk_-}\,|\,P\,)-8\de P\log(R/2\pi)\,\in\,\pi \BZ\,,
\end{equation}
where $\theta_\pm (\,{\bk_\pm}\,|\,P\,)$ 
are the phases of the eigenvalues of ${\mathbf T}^{\pm}(P)$ as introduced
above. Introducing a positive integer $m$ which labels
the solutions $P_{m}$ of \rf{Liouquant} we are lead to the conclusion 
that the UV asymptotics for the spectrum of the Sinh-Gordon model 
is given by
\begin{equation}\label{UVenergy2}
E(\bk_+,\bk_-,m)
\,=\,\frac{2\pi}{R}\bigg( 2P_{m}^2-\frac{1}{12}+\sum_{a\in \BK_+}
k_a+\sum_{a\in \BK_-}k_a\bigg)\,,
\end{equation}
with $P_{m}=P_{m}(\bk_+,\bk_-)$ being a solution to \rf{Liouquant}.
We observe a remarkable agreement with the 
description of $\CH_{\rm TBA}$ that we had found in 
subsection \ref{UVspec1}. 

Comparing formulae \rf{Thetadef} and \rf{refamp} 
we furthermore observe that the quantization conditions
\rf{Liouquant} and \rf{Q-cond} precisely agree in the 
case $\BK_\pm=\emptyset$ 
provided that the parameters 
$m$ and $\mu$ are related as
\begin{equation}\label{m-mu}
\frac{\pi\mu}{\ga(-b^2)}\,=\,-\left[\frac{m}{4\sqrt{\pi}}\,
\Ga\bigg(\frac{1}{2+2b^2}\bigg)
\Ga\bigg(1+\frac{b^2}{2+2b^2}\bigg)\right]^{2+2b^2}\,.
\end{equation}

These  observations not only represent a highly
nontrivial confirmation of the picture proposed
in this paper, they also lead
to a re-derivation of the
important result from \cite{Z3} concerning the exact relation 
between the scale parameters $\mu$ and $m$
defined in terms of the
UV and the IR behavior of the Sinh-Gordon model, 
respectively.

On the other hand we would like to point out that our discussion above
leads us to conjecture that 
\begin{equation}
2\Theta(\,\bk_\pm|\,P\,)\,=\,T(P)+2\theta_\pm (\,{\bk_\pm}\,|\,P\,)\,.
\end{equation}
This relation would represent a remarkable and highly 
nontrivial link between 
the integrable structure of quantum KdV theory with $c>1$ 
- as encoded in $\Theta(\,\bk_\pm|\,P\,)$ -  and the 
conformal structure of quantum Liouville theory that was
used in the definition of $\theta_\pm (\,{\bk_\pm}\,|\,P\,)$,
respectively.


\section{Concluding remarks}
\setcounter{equation}{0}

The completeness of  $\CH_{\rm\sst TBA}$ within
$\CH_{\rm\sst SG}$ (Conjecture \ref{complconj})
may further be checked by 
an analysis of the semiclassical limit $b\ra 0$. 
This is currently under investigation. Our preliminary results
seem to support Conjecture \ref{complconj}. 

It seems natural to compare our results to the description 
of the Sine-Gordon spectrum in terms of nonlinear
integral equations (NLIE) that was found by Destri and De Vega in  
\cite{DDV1,DDV2,DDV3}, see also \cite{FRT}.
A quick comparison reveals important differences between
the two cases, however. We note, in particular, that the functions
constrained by the respective nonlinear integral equations 
are related to the Q-functions in different ways, as can be
inferred from \cite{BLZ0,BLZ2}. In the case of the Sinh-Gordon
model one would have to look for a nonlinear integral equation 
satisfied by the function
\begin{equation}
a(u)\,=\,\frac{q(u+ib)}{q(u-ib)}\,,
\end{equation}
in order to be able to compare the NLIE of Destri-De Vega type more
directly to the NLIE which we have discussed here.
It should be very illuminating to analyze in some detail how the
physical differences between the two models are encoded in the 
analytic structure of the $Q$-functions that characterize the 
respective spectra.

\newpage
\appendix
\section{Derivation of the nonlinear integral equations}
\setcounter{equation}{0}

Our aim will be to show that the elements of $\CQ_M$
are in one-to-one correspondence 
to the solutions of certain nonlinear integral equations.

\subsection{Renormalising the Q-functions}

The first step is to introduce a conveniently renormalised Q-function $Q(u)$, 
\begin{equation}\label{Qfromq}
Q(u)\,=\,\Lambda_{\SRN}(u)q(u)\,,\qquad
\Lambda_{\SRN}(u):=\big[\Phi_b(-s+u-i\de)\Phi_b(-s-u-i\de)\big]^{\SRN}\,,
\end{equation}
with special function $\Phi_b(u)$ defined as
\begin{equation}
\Phi_b(u):=\exp\left(\int_{\BR+i0}\frac{dt}{8t}\;\frac{e^{-2itu}}{\sinh(bt)\sinh(t/b)
\cosh(2t\de)}\right)\,.
\end{equation}
The special function $\Phi_b(u)$ was introduced in \cite{BMS}, where the following 
properties were listed:
\begin{itemize}
\item[(i)] Functional relation
\begin{equation}\label{Phifunrel}
\Phi_b(u+i\de)\Phi_b(u-i\de)\,=\,e_b(u)\,,
\end{equation}
\item[(ii)] Asymptotic behavior 
\begin{equation}
\Phi_b(u)\,\sim\,\left\{
\begin{aligned} 1\qquad\quad & \;\;{\rm for}\;\;|\arg(- u)|<\frac{\pi}{2}\\
e^{\mathrm{i}\frac{\pi }{2}u^2-\mathrm{i}\frac{\pi }{12}(1-8\de^2)} &   
\;\;{\rm for}\;\; |\arg(+ u)|<\frac{\pi}{2}\,.
\end{aligned}\right.
\end{equation}
\item[(iii)] Zeros and poles:
\[
\begin{aligned}
&\text{poles at}\;\,z=+\mathrm{i}(2\de+mb+nb^{-1}),\quad m,n\in\BZ_{\geq 0},\quad m+n-|m-n|=0\;\textrm{mod}\; 4,\\
&\text{zeros at}\;\,z=-\mathrm{i}(2\de+mb+nb^{-1}),\quad m,n\in\BZ_{\geq 0},\quad m+n-|m-n|=0\;\textrm{mod}\; 4.
\end{aligned}
\]
\end{itemize}
It follows from \rf{Phifunrel}
that 
\begin{equation}
W_{\SRN}(u)=(\Lambda_{\SRN}(u+i\de)\Lambda_{\SRN}(u-i\de))^{-1},
\end{equation} implying that
\rf{Wron} is equivalent to 
\begin{equation}\label{Wr'}
Q(u+i\de) \, 
Q(u-i\de)\,=\,1+\frac{1}{\Xi_{\SRN}(u)}Q(u+i\de')Q(u-i\de')
\,.
\end{equation}
with
\begin{equation}\label{Xisimple}
\Xi_{\SRN}(u)\,=\,\frac{\Lambda_{\SRN}(u+i\de')
\Lambda_{\SRN}(u-i\de')}{\Lambda_{\SRN}(u+i\de)\Lambda_{\SRN}(u-i\de)}
\,=\,\bigg(\frac{\cosh\frac{\pi}{2\de}s+\cosh\frac{\pi}{2\de}u}
             {\cosh\frac{\pi}{2\de}s-\cosh\frac{\pi}{2\de}u}\bigg)^N\,.
\end{equation}

The relevant analytic properties of $Q(u)$ defined in \rf{Qfromq} are found
from the properties of $q(u)$ listed above together with the properties (ii) and (iii) of $\Phi(u)$:
{\it \begin{itemize}
\item[(a)]
$Q(u)$ is asymptotic to a constant, 
\begin{equation}\label{Q-asym}
Q(u)\,\sim\,\frac{1}{\sqrt{2}}\,,\qquad |u|\ra\infty\,,\;\;|\arg(\pm u)|<\frac{\pi}{2}\,,
\end{equation}
\item[(b)] 
$Q(u)$ is analytic in the strip 
$\BS_\ep\equiv\{\,z\in\BC\,;\,|{\rm Im}(z)|< \de+\ep\,\}$ if $\ep<b^{\pm 1}$.
\end{itemize}}
The definition \rf{Qfromq} removes the poles of $q(u)$ at $u=\pm(s-\mathrm{i}\de)$. 
The new poles introduced 
in this way have distance at least $2\mathrm{i}\de$ from the real axis.
Note furthermore that the limiting value of $Q(u)$ in \rf{Q-asym} is easily determined from 
\rf{Wr'} noting that 
$\Xi_\SRN(u)$ is asymptotic to $-1$ for odd $\SRN$. It follows 
that the constant $Q_0$ in the asymptotics 
\rf{Q-asym} is equal to $Q_0=1/\sqrt{2}$.

\subsection{Reformulation in terms of integral equations}

The next step is to define $Y(u)$ by
\begin{equation}\label{Ydef}
Y(u):=Q(u+i\de)Q(u-i\de)-1\,.
\end{equation}
The function $Y(u)$ is analytic in a strip of width smaller than $\min(b^{\pm 1})$
around the real axis, and asymptotic to a constant when $|\mathrm{Re}(u)|\ra \infty$,
up to corrections that decay exponentially.

One may reconstruct $Q(u)$ from $Y(u)$ by writing \rf{Ydef} as
\begin{equation}\label{Xdef}
\log(Q(u+i\de))+\log (Q(u-i\de))\,=\,\log\left(1+Y(u)\right),
\end{equation}
and inverting the 
difference operator on the left hand side. Care is needed when 
$Q(u)$ has zeros in $\BS$. The resulting formula looks as follows: 
\begin{equation}\label{Qreconst}
\log(Q(u))=
\int_\BR\frac{dv}{2\de}\;\frac{1}{2\cosh\frac{\pi}{2\de}(u-v)}
\log\left(1+Y(v)\right)+
\sum_{a=1}^M\log\tanh\frac{\pi}{4\de}(u-u_a)\,.
\end{equation}
In order to see this it is convenient to differentiate \rf{Qreconst} with respect to the variable $u$, 
\begin{equation}\label{Qreconst'}
\frac{Q'(u)}{Q(u)}=
\int_\BR\frac{dv}{2\de}\;\frac{1}{2\cosh\frac{\pi}{2\de}(u-v)}
\frac{\pa}{\pa v}\log\left(1+Y(v)\right)+
\sum_{a=1}^M\frac{\pi}{2\de \sinh\frac{\pi}{2\de}(u-u_a)}\,.
\end{equation}
The integral on the left of \rf{Qreconst'} is a convolution of meromorphic functions which is analytic inside
of strips $\BS_e$ with $\ep<\min(b^{\pm 1})$. 
The zeros of $Q(u)$ are displayed explicitly 
in the  second term of the right of \rf{Qreconst'}. It is straightforward
to check that the function $\frac{\pa}{\pa u}\log Q(u)$ defined via \rf{Qreconst'} 
satisfies the equation obtained by differentiating \rf{Xdef}. It follows that $Q(u)$ defined in \rf{Qreconst}
satisfies \rf{Xdef} up to a constant which is found by considering the limit $u\ra\infty$. 

The quantum Wronskian relation \rf{Wr'} implies that $\log Y(u)$ can alternatively
be represented as
\begin{equation}\label{QWr'}
\log Y(u)\,=\, \log Q(u+i\de')+\log Q(u-i\de')-
\log\Xi_n(u)\,.
\end{equation}
It follows from \rf{QWr'} that 
\begin{equation}
\log\left(1+Y(v)\right)\,\sim\,\log\left(1-\frac{1}{2}\right)\,,\quad
{\rm for}\;\;|v|\ra\infty\,,
\end{equation}
using that $\Xi_{\SRN}(v)\sim -1$ for odd $\SRN$ together with the asymptotics of $Q(u)$.
Noting furthermore
\begin{equation}
\int_\BR\frac{dv}{2\de}\;\frac{1}{2\cosh\frac{\pi}{2\de}(u-v)}\,=\,
\frac{1}{2\pi}\int_{\BR}\frac{dx}{\cosh(x)}\,=\,\frac{1}{2}\,.
\end{equation}
it follows that the expression for $\log Q(u)$ given in \rf{Qreconst} is asymptotic to 
$
-\log\sqrt{2}=\log Q_0
$, as it should be.

Inserting the expression for $\log(Q(u))$ from \rf{Qreconst} into the right hand side of \rf{QWr'}
yields the following nonlinear integral equation
\begin{equation}
\label{NLIE}
\begin{aligned}
\log  Y(u) &= \,\int_{{\BR}}
\frac{dv}{4\de}\;\si(u-v) \log\left(1+Y(v)\right)\\
&\qquad-
\log\Xi_\SRN(u)-\sum_{a=1}^{\SRM} \log S(u-u_a-i\de)\,,
\end{aligned}
\end{equation}
using the following notations
\begin{align}
\si(u)\,&\,=\,
\frac{1}{\cosh\frac{\pi}{2\de}(u-i\de')}+
\frac{1}{\cosh\frac{\pi}{2\de}(u+i\de')}\,,
\\
S(u-i\de)&=\frac{1}{\tanh\frac{\pi }{4\de}(u+i\de-ib)\tanh\frac{\pi }{4\de}(u-i\de+ib)}\,.
\label{S-def}\end{align}
The function $S(u)$ defined in \rf{S-def} can be rewritten in more familiar form as
\begin{equation}
S(u)\,=\,
\frac{\tanh\frac{\pi }{4\de}(u-ib)}{\tanh\frac{\pi }{4\de}(u+ib)}=  \frac{\sinh\frac{\pi }{2\de}u-i\sin\theta_0}
      {\sinh\frac{\pi }{2\de}u+i\sin\theta_0}\notag\,.
      \end{equation}
Equation \rf{NLIE} uniquely defines functions $Y_{\mathbf u}(u)$ depending parametrically on the 
input data ${\mathbf u}=(u_1,\dots,u_M)$. From these functions one may  reconstruct the
functions $Q(u)$ via \rf{Qreconst}. 

However, a necessary condition for the resulting functions to satisfy
\rf{Wron} is that 
\begin{equation}
1+{Y(u_a+i\de)}\,=\,0 \quad\text{for} \quad a=1,\dots,M,
\end{equation}  
which is equivalent to 
\begin{equation}\label{qcond}
\begin{aligned}
\pi(2k_a+1)\,=&\,\int_{{\BR}}
\frac{dv}{2\de}\;\tau(u_a-v) \log\left(1+{Y_{\mathbf u}(v)}\right)\\[-.5ex]
&\qquad
+\sum_{\substack{b=1\\ b\neq a}}^{\SRM} \arg S(u_a-u_b)
+2\SRN\arctan\left(\frac{\sinh\frac{\pi}{2\de}u_a}{\cosh\frac{\pi}{2\de}s}\right),\end{aligned}
\end{equation}
where 
\begin{equation}
\tau(u)=i\si(u+i\de)=\frac{4\sin\frac{\pi }{2\de}b\,\sinh\frac{\pi}{2\de}u}
{\cosh\frac{\pi}{\de}u+\cos\frac{\pi }{\de}b}\,.
\end{equation}
This is a system of nonlinear equations restricting the choice of the data ${\mathbf u}=(u_1,\dots,u_M)$
severely. 

\section{Proof of Proposition \ref{confdim}}
\setcounter{equation}{0}

To begin with, let us note that summing over $a$ in the
relations (B$_{\pm}$) yields the relations
\[
\begin{aligned}
\pm\frac{mR}{2}\sum_{a\in\BK_\pm} e^{\pm\vartheta_a}
-
\int_{\BR}\frac{d\vartheta}{2\pi}\,\log(1+Y_{{\bk_\pm}}^\pm(\vartheta))
\sum_{a\in\BK_\pm}\tau(\vartheta_a-\vartheta) \,=\,
2\pi\sum_{a\in\BK_\pm}k_a\,.
\end{aligned}
\]
Inserting the expressions for $\sum_{a\in\BK_\pm}
\tau(\vartheta_a-\vartheta)$ which 
follow from (I$_\pm$)
leads to the formula
\begin{align}\label{level+I}
& \pm\frac{mR}{2}\Bigg(\sum_{a\in\BK_\pm} e^{\pm\vartheta_a}  -\int_{\BR}
\frac{d\vt}{\pi}\,e^{\pm\vt}\log(1+Y^{\pm}_{{\bk_\pm}}(\vt))\Bigg)=
2\pi\sum_{a\in\BK_\pm}k_a+\CI_{{\bk_\pm}}^\pm
\end{align}
where $\CI^{\pm}_{\bk_\pm}$ is defined as
\begin{align}
\CI_{{\bk_\pm}}^\pm\,\equiv\,
\int_{\BR}
\frac{d\vt}{2\pi}\,\log(1+Y^{\pm}_{{\bk_\pm}}(\vt))\frac{\pa}{\pa\vt}\bigg(
\log Y^{\pm}_{{\bk_\pm}}(\vt)-\int_{\BR}
\frac{d\vt'}{2\pi}\,\si(\vt-\vt')\log(1+Y^{\pm}_{{\bk_\pm}}(\vt'))\bigg)\,.
\end{align}
This integral can be evaluated explicitly:
\begin{propn}
\begin{align}\label{CIcalc}
\CI_{{\bk_\pm}}^{\pm}\,=\,\pm 2\pi\Big(P^2-\frac{1}{24}\Big)\,.
\end{align}\end{propn}
\begin{proof}
We note that the integrals defining $\CI^{\pm}_{\bk_\pm}$ are absolutely 
convergent. We may therefore perform the
calculation under the assumption that $\Im(P)>0$ and take the
limit $\Im(P)\ra 0$ in the end.

Let us first consider 
\begin{equation}\label{FFint}
\int_{-\xi}^\xi\frac{d\vt}{2\pi}\int_{-\zeta}^\zeta\frac{d\vt'}{2\pi}
\,F(\vt)\,\frac{\pa}{\pa\vt}\si(\vt-\vt')\,F(\vt')
\,,
\end{equation}
where $F(\vt)\equiv\log(1+Y^{+}_{\bk_+}(\vt))$.
The anti-symmetry of the kernel $\frac{\pa}{\pa\vt}\si(\vt-\vt')$ in
\rf{FFint} implies that the integral \rf{FFint} can be written as
\begin{equation}\label{FFint2}
\int_{-\xi}^\xi\frac{d\vt}{2\pi}\;F(\vt)\bigg(
\int_{-\zeta}^{-\xi}\frac{d\vt'}{2\pi}
\,\frac{\pa}{\pa\vt}\si(\vt-\vt')\,F(\vt')+
\int_{\xi}^{\zeta}\frac{d\vt'}{2\pi}
\,\frac{\pa}{\pa\vt}\si(\vt-\vt')\,F(\vt')\bigg)
\end{equation}
We will ultimately send $\xi,\zeta\ra\infty$ such that $0<\xi<\zeta$.
This allows us to approximate $F(\vt')$ in \rf{FFint2} by its 
asymptotics
\begin{equation}\label{Fasy}
F(\vt')\,\sim\,\left\{\begin{aligned} 0\;\;&\text{for}\;\;\vt'\ra\infty\,,\\
8i\de P\vt'\;\;&\text{for}\;\;\vt'\ra -\infty\,,
\end{aligned}\right.
\end{equation}
which leads us to represent the integral in \rf{FFint2} as
\begin{equation}
\frac{8i\de P}{2\pi}\int_{-\xi}^\xi\frac{d\vt}{2\pi}\,F(\vt)
\Big(\xi\si(\vt+\xi)-\zeta\si(\vt+\zeta)-(
\arg S(\vt+\xi)-\arg S(\vt+\zeta))\Big)\,.
\end{equation}
Noting that $\si(\vt)$ and $\arg S(\vt)$ are peaked around
$\vt=0$ allows us to use the asymptotics  \rf{Fasy} for 
$F(\vt)$, leading to us to approximate \rf{FFint2} by
\begin{equation}\label{FFapprox}
\frac{(8\de P)^2}{8\pi^2}\int_{\BR}{d\vt}\,\Big(\xi^2\si(\vt)
-\frac{1}{2}\vt^2\si(\vt)\Big)\,.
\end{equation}
\begin{lem}\label{intlem}
\[
\int_\BR d\vt\,\si(\vt)=\,2\pi\,,\qquad
\int_\BR d\vt\,\vt^2\,\si(\vt)\,=\,2\pi\frac{\pi^2}{(2\de)^2}\,.
\]
\end{lem}
\begin{proof}
In both cases we may start by shifting the contour of integration into the upper half plane. 
The integrals above are equal to the sum of integrals over $\BR+i\pi$ plus
the sum of residues of poles the integrands have between $\BR$ and $\BR+i\pi$. 
The integrals over $\BR+i\pi$ may be expressed in terms of integrals above due to $\si(\vt+\pi)=-\si(\vt)$.
Taking into account that $\si(\vt)=\si(-\vt)$ we easily reduce the evaluation of the integrals to 
the computation of the sums of the residues.
\end{proof}
It follows from \rf{FFapprox} together
with Lemma \ref{intlem} that the leading asymptotics of \rf{FFint2}
for $\xi\ra\infty$ is given by the terms
\begin{equation}\label{FFapprox2}
\frac{(4\de P)^2}{\pi}\xi^2-2\pi P^2\,.
\end{equation}

It remains to consider the first term in the definition of 
$\CI^{+}_{\bk_+}$, regularized as
\begin{align}\nn
\int\limits_{-\xi}^\xi\frac{d\vt}{2\pi}\,
& \log(1+Y^{+}_{\bk_+}(\vt))\, \frac{\pa}{\pa\vt}\log
Y^{+}_{{\bk_+}}(\vt)\,=\!\!\!\int\limits_{Y^{+}_{\bk_+}(-\xi)}^{Y^{+}_{\bk_+}(\xi)}
\frac{dY}{2\pi}\,\frac{\log(1+Y)}{Y} \\
&
\,=-\int\limits^{\log Y^{+}_{\bk_+}(-\xi)}_{0}
\frac{dx}{2\pi}\,(x+\log(1+e^{-x}))-
\int\limits^{0}_{\log Y^{+}_{\bk_+}(\xi)}
\frac{dx}{2\pi}\,\log(1+e^{x})\label{firstterm}
\,.
\end{align}
In the limit when $\xi\ra\infty$ we may note that \rf{Fasy} implies that
$\log Y^{+}_{\bk_+}(\xi)\sim\log F(\vt)\ra -\infty$, 
whereas $\log Y^{+}_{\bk_+}(-\xi)\sim -8i\de P\xi$. We therefore 
may estimate the leading behavior of \rf{firstterm} as
\begin{equation}\label{firstapprox}
\int\limits_{-\xi}^\xi\frac{d\vt}{2\pi}\,
\log(1+Y^{+}_{\bk_+}(\vt))\, \frac{\pa}{\pa\vt}\log
Y^{+}_{{\bk_+}}(\vt)\,\sim\,\frac{1}{2}\,\frac{(8\de P)^2}{2\pi}\xi^2-
2\int\limits_0^1\frac{dY}{2\pi}\,
\frac{\log(1+Y)}{Y}\,.
\end{equation}
Subtracting \rf{FFapprox2} from \rf{firstapprox}
and using $\int_0^1\frac{dY}{2\pi}\,
\frac{\log(1+Y)}{Y}=\frac{\pi}{24}$ concludes the proof
of the Proposition for the case of $\CI^{+}_{\bk_+}$. The
case of $\CI^{-}_{\bk_-}$ can be treated in a similar way. 
\end{proof}
Inserting formula \rf{CIcalc} 
into \rf{level+I} concludes the proof of Proposition \ref{confdim}.

\end{document}